\documentclass[aip,reprint]{revtex4-1}
\usepackage{amsmath}
\usepackage{amssymb}
\usepackage{graphicx}% Include figure files
\usepackage{dcolumn}% Align table columns on decimal point
\usepackage{bm}% bold math
\usepackage{multirow}
\usepackage{hyperref}

\begin{document}

% \title{Brownian Dynamics driven assembly of patch particles:\\Effects of Rotational Degeneracy}% 
\title{Controlling morphology in hybrid isotropic/patchy particle assemblies}

\author{Srinivas Mushnoori}
\affiliation{Department of Chemical and Biochemical Engineering, Rutgers, The State University of New Jersey, Piscataway, New Jersey 08854}%Lines break automatically or can be forced with \\
\author{Jack A. Logan}
\affiliation{Department of Physics and Astronomy, 
Stony Brook University, Stony Brook, NY 11794}
\affiliation{Center for Functional Nanomaterials, 
Brookhaven National Laboratory, Upton NY 11973}

\author{Alexei V. Tkachenko}
\email{oleksiyt@bnl.gov}
\affiliation{Center for Functional Nanomaterials, 
Brookhaven National Laboratory, Upton NY 11973}

\author{Meenakshi Dutt }
\email{meenakshi.dutt@rutgers.edu}
\affiliation{Department of Chemical and Biochemical Engineering, Rutgers, The State University of New Jersey, Piscataway, New Jersey 08854}

\date{\today}

\begin{abstract}

Brownian Dynamics is used to study self-assembly in a hybrid system of istotropic particles (IPs), combined with anisotropic building blocks that represent special ``designer particles".  Those are  modeled as spherical patchy particles (PPs) with binding only allowed between their patches and IPs. In this study, two types of PPs are considered: Octahedral PPs (Oh-PPs) and Square PPs (Sq-PPs), with octahedral and square arrangements of  patches, respectively.  The self-assembly is additionally facilitated by the simulated annealing procedure. The resultant structures are characterized  by a combination of local correlations in  cubatic  ordering, and a symmetry-specific variation  of bond orientation order parameters (SymBOPs). By varying the PP/IP size ratio, we detected a sharp crossover between two distinct morphologies, in both types of systems. High symmetry phases, NaCl crystal for Oh-PP and square lattice for Sq-PP, are observed for larger size ratios. For smaller ones, the dominant morphologies are significantly different, e.g., Oh-PPs form a compact amorphous structure with predominantly Face-to-Face orientation of neighboring PPs. Unusually for a morphology without a long range order, it is still possible to identify well organized coherent clusters of this structure, thanks to the adoption of our SymBOP-based characterization.

% Shape controllable aggregation of particles is an interesting domain that allows the creation of spontaneously assembled large scale geometries. We propose a model of octahedral DNA structures that aggregate into well defined lattices. A coarse-grained model is developed and utilized to test two systems: one with completely symmetric octahedra, one with asymmetric octahedra.  
% \begin{description}
 
% \end{description}
\end{abstract}

\maketitle

\section{\label{sec:level1}Introduction}
%https://www.overleaf.com/project/5f848ecb8611f70001fd7097
The increasing prominence of the field of programmable self-assembly is rooted in its potential to revolutionize nano- and micromanufacturing in diverse disciplines such as photonics, plasmonics or medicine. The  key idea is to use specially built ``designer particles" capable of assembly into desired structures, such as superlattices or mesoscopic shapes \cite{Glotzer_Solomon2007, mjstevens_nanotubes}. Examples of such designer building blocks are DNA-functionalized micro- and nanoscale isotropic particles \cite{Gang_DNA,Mirkin_DNA,Macfalane,Pine2015, Tkachenko2002,Francisco, Vini_review, Dijks2014}, as well as  ``patchy" colloids featuring a pattern of chemically distinct spots \cite{Pine_patchy_2003,Pine_2012,Patchy_fab,patchy_stefano_2013,Diamond_2020}. The former approach allows the control of interactions between  individual particles by using DNA hybridization. Whereas the latter approach involves anisotropic interparticle interactions, allowing for control over the symmetry of the desired self-assembled nano- and microstructure. More recently, an amalgamation of the two approaches has been adopted to give rise to building blocks based on DNA origami, which combine key-lock interactions with anisotropy \cite{Gang_Frames,Oleg_Octahedra_2020,John_PRE2013}. Furthermore, it has been demonstrated that by combining these origami-based designer building blocks, with more conventional  isotropic DNA-functionalized nanoparticles, one can achieve remarkable control over the symmetry and overall  architecture of the assembly of the system components \cite{tian2015prescribed,Oleg_hybrid_2016}. In particular, such an approach has been utilized to self-assemble a nanoscale diamond lattice \cite{diamond_oleg}, long considered the ``Holy Grail" of the field of self-assembly \cite{Diamond_2020}. In this study, self-assembled nanostructures in hybrid systems encompassing both patchy and isotropic particles are explored using molecular simulations. The study is relevant both for microscale colloids, and nanoscopic building blocks, where   the patchy particles represent multivalent building blocks based on DNA origami.         

Numerous computational studies have been undertaken to better understand the mechanistic aspects of assembly of patchy particles and their emergent self-assembled structures. Their phase behavior and kinetics  have been studied extensively using Monte Carlo  and Molecular Dynamics (MD)  methods \cite{phase_diagram_patchy,Sciort_patchy_PRL,Glo04,bianchi_crystal,Bianchi_phases,Vasilyev_1,Vasilyev_chrom, Patra2018,JCP2020_Sharon}. Majority of those   studies have discussed structures obtained  by direct binding between anisotropic particles.  In this work, inspired by recent experiments \cite{tian2015prescribed,diamond_oleg,Oleg_hybrid_2016},   we consider a hybrid case when patchy particles (PPs) do not directly bind with each other, but instead are capable of binding to isotropic particles (IPs). In addition, motivated by the experimental reality, we implement an annealing scheme in which the binding strength is temporally modulated  during the self-assembly  process. Finally, the results are analyzed by using a new characterization method which generalizes the widely used concept of orientational bond order parameters, by taking advantage of the symmetry of constituent building blocks, i.e., PPs. Specifically, we use Symmetrized Bond Order Parameters (SymBOPs) introduced in Ref.~\onlinecite{symbop} to extract the structural information about the system.

\section{\label{sec:level1}Methods}

\subsection{\label{sec:model} Model}
\begin{figure}[h!]
    \centering
    \includegraphics[width=0.8\columnwidth]{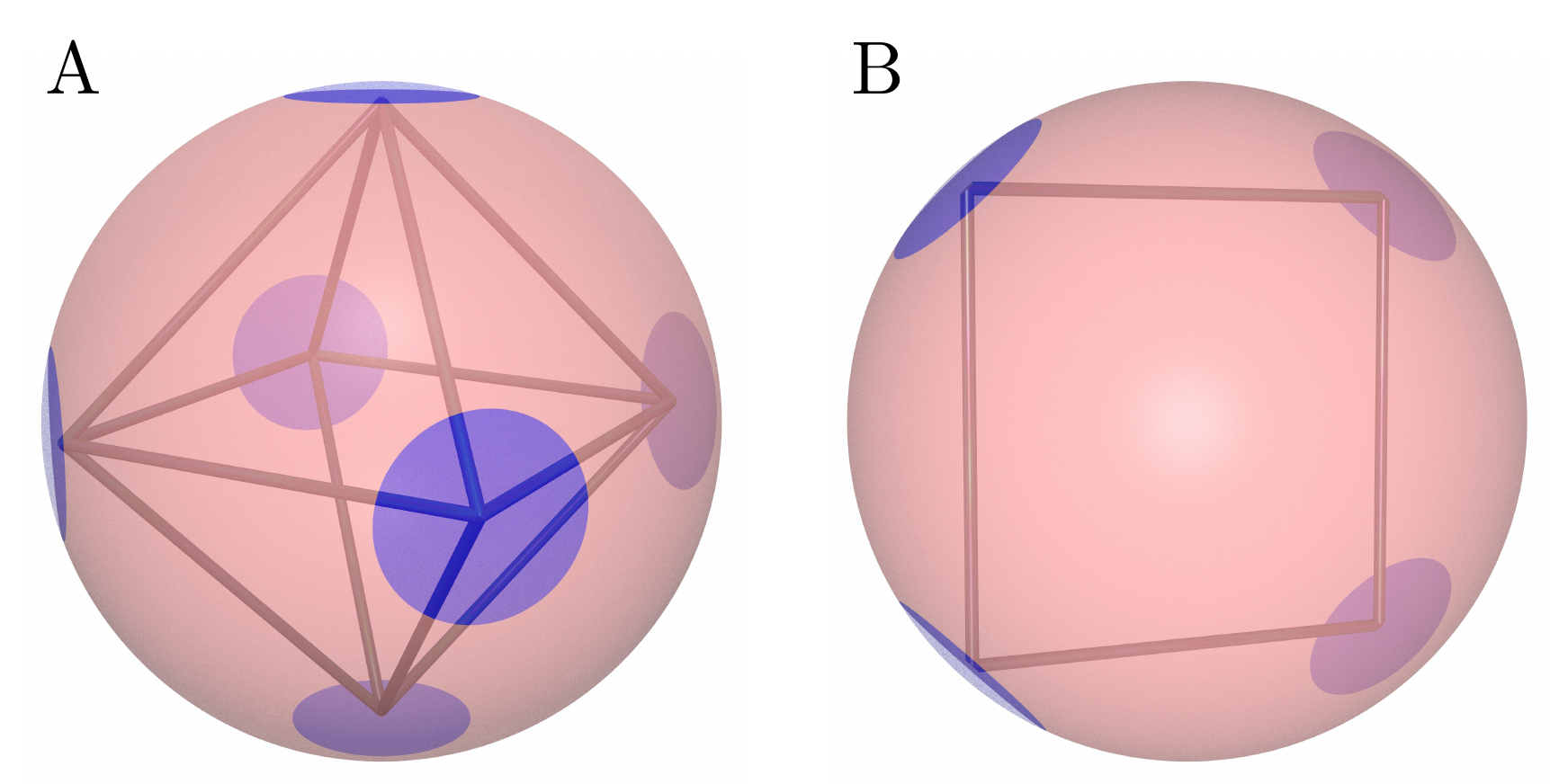}
    \caption{Schematic diagrams of the (A) Oh-PPs and (B) Sq-PPs. Blue regions represent the interactive patches, located at the vertices of the particles. The central pink sphere is the central repulsive core. For clarity, the central repulsive core has been made translucent to show the positions of the vertices on the patches.}
    \label{fig:patchy_particles}
\end{figure}

The self-assembly in a hybrid system of IPs and PPs  is investigated by means of Brownian Dynamics (BD) simulations. In our coarse grained simulations, both types of particles are modeled as hard spheres of diameters $\sigma_{IP}$ and  $\sigma_{PP}\equiv \sigma$, respectively.  Their centers are represented as vectors ${\bf R}^{(\alpha)}_{i}$, where $\alpha \in\{IP,PP\}$. The ``patches" are interaction sites that reside at specific locations on the surface of PPs to enable anisotropic interactions. Their locations  are represented as vectors ${\bf r}^{(n)}_i$, where  $i$ is an index of the respective PP, and $n=1...N_p$ is the patch index. Each PP, together with its  patches are modeled as a rigid body.    In this study, two designs  of PPs are considered, featuring octahedral ($N_p=6$) and square ($N_p=4$) arrangements of patches, shown in Fig.~\ref{fig:patchy_particles}.   We will refer to them as to Oh-PPs, and Sq-PPs,  respectively.      

\begin{figure}[h!]
    \centering
    \includegraphics[width=0.7\columnwidth]{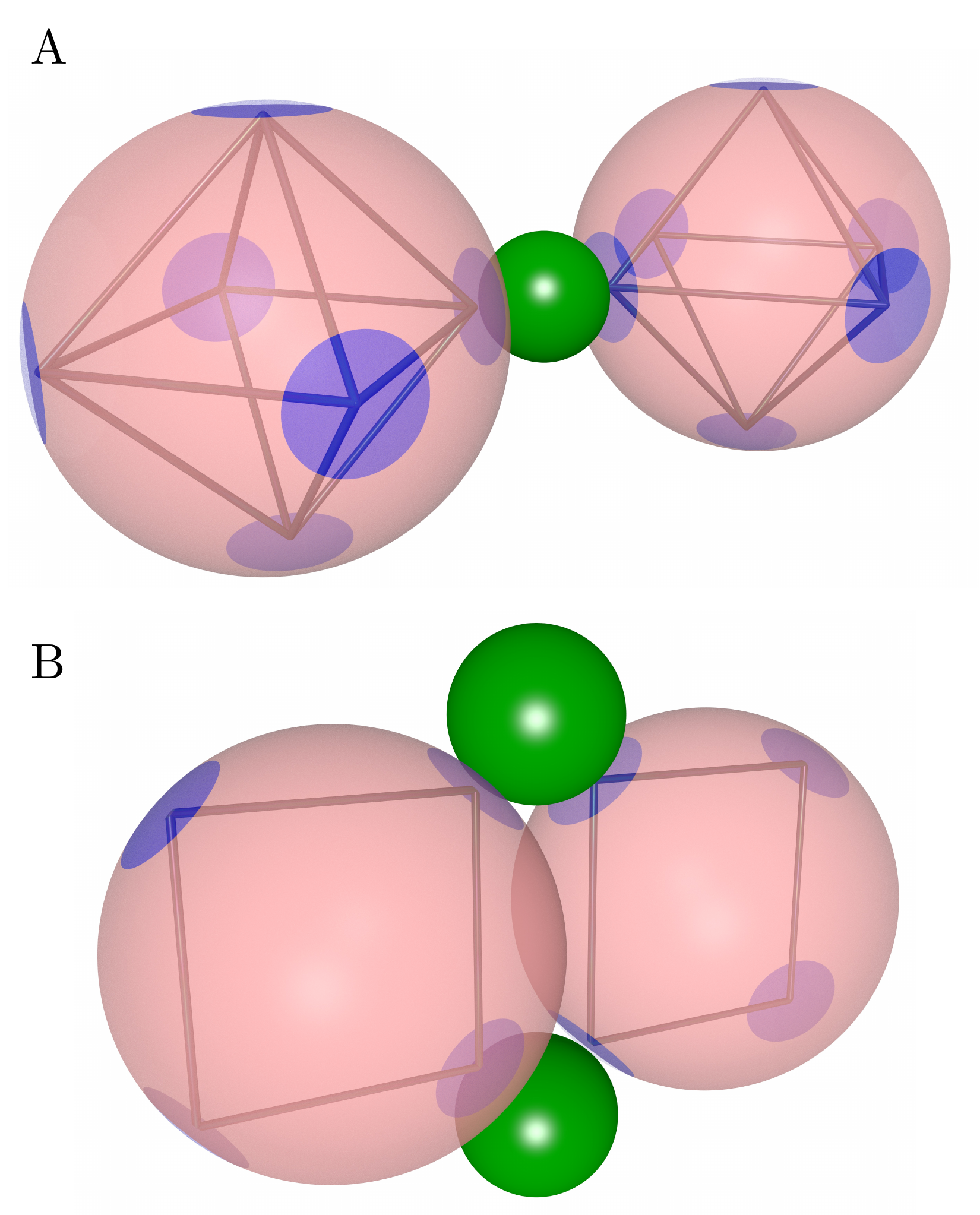}
    \caption{Bonding between PPs is facilitated by spherically symmetric IPs. (A) Two Oh-PPs  bind vertex-to-vertex via one IP. (B) Two Sq-PPs bind edge-to-edge via two IPs.}
    \label{fig:PP_bonding}
\end{figure}

The hard-core repulsion between any pair of particles is implemented in the form of  Weeks-Chandler-Anderson (WCA) repulsive  with a hard core parameter  $\sigma_{\alpha\beta}=\sigma_{\alpha}+\sigma_{\beta}$ (here $\alpha,\beta\in \{IP,PP\}$:
\begin{equation} \label{eq:wca_pot}
V_{\alpha\beta}(r) = 
\begin{cases}
4\epsilon \left[(\frac{\sigma_{\alpha\beta}}{r})^{12} - (\frac{\sigma_{\alpha\beta}}{r})^6\right] + \epsilon, r < 2^{1/6}\sigma_{\alpha\beta} \\
0                     , r \ge 2^{1/6}\sigma_{\alpha\beta}
\end{cases}
\end{equation}
In our study,  PPs do not  bind to each other directly. A soft  attractive potential between  patches and IPs is given by
\begin{equation} \label{eq:soft_pot}
V_{p}(r) = 
\begin{cases}
-\epsilon[1 + \cos( \pi \frac{r}{R_c})], r < R_c \\
0                     , r \ge R_c
\end{cases}
\end{equation}
Here $R_c=0.5\sigma$,  $r$ is the patch-to-IP distance. In the model, $\sigma$ and $\epsilon$ are the fundamental length and energy scales, respectively. The total interaction potential thus has the form  
\begin{equation} \label{eq:total}
U_{\rm tot} = \sum_{i,j,\alpha,\beta} V_{\alpha \beta}\left(\left|{\bf R}^{(\alpha)}_i-{\bf R}^{(\beta)}_j\right|\right)+ \sum_{i,j,n}{ V_{p}\left(\left|{\bf R}^{(IP)}_i +{\bf r}^{(n)}_{j}\right|\right)} \\
 \end{equation}
 Here $\alpha,\beta\in \{IP,PP\}$, $n=1..N_p$, and indices $i$ and $j$ run over all particles of a given type, without double-counting. Figure~\ref{fig:PP_bonding} shows  examples of binding between two PPs  facilitated by the bridging IPs.

% The organization of the particles within the assemblies is characterized by a bond order parameter discussed in the Characterization section.

BD simulations are performed via the open source community-based MD package Large-scale Atomic/Molecular Massively Parallel Simulator (LAMMPS). The simulations are performed using simulations boxes with three dimensional periodic boundary conditions and sample the canonical ensemble (NVT).
%The timestep is set to 0.001$\tau$.  
For the initial set up,  a preliminary simulation cell of dimension $1.5~\sigma$ encompassing a single PP and a single IP is created (i.e., the system has 1:1 composition). The cell is replicated in the X, Y and Z dimensions 8-10 times to generate simulation boxes with dimensions  12-15~$\sigma$ with a total of 512-1000 PPs (as well as the corresponding IPs). The resultant simulation boxes are thermally spiked with soft repulsive potentials active for 10,000 time steps to induce randomness in the spatial configuration of the particles. This step is followed by self assembly, facilitated by the annealing. 

\subsection{\label{sec:annealing} Annealing}

\begin{centering}
\begin{figure}[h]
\includegraphics[width=1\columnwidth]{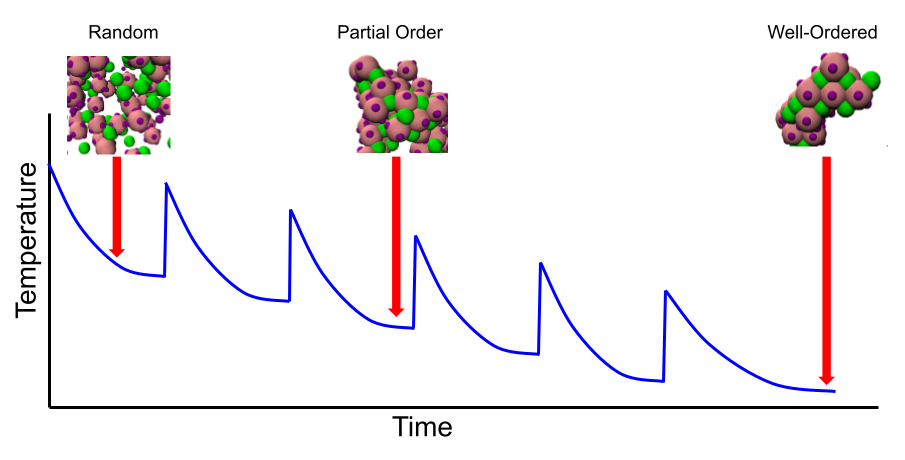}
\caption{
% \textbf{REWORK to use similar scheme for VMDs as Jack's images} 
The time evolution of the temperature is shown along with insets of the local structure after each phase. Inset 1 (left) shows disordered aggregation. Inset 2 (middle) shows partially ordered assembly. Inset 3 (right) shows a well-ordered assembly. The annealing process allows the system to first form a randomly aggregated mass, after which it is thermally spiked and slowly allowed to cool in order to thermodynamically relax unfavourable bonds. The final structure is a well-relaxed, ordered structure. }
\label{fig:annealing}
\end{figure}
\end{centering}

Systems that are expected to yield highly ordered structures rarely  do so via direct or spontaneous assembly. This is largely due to kinetic limitations and tendency towards glassy behavior.  To address this, real-world  self-assembly is often subjected to  annealing procedures. This allows the systems to escape  disordered metastable  configurations and find new, more favorable morphologies. In our study, we likewise implement the annealing protocol. Specifically, the system is first allowed to run for 250,000 time steps to promote its collapse into a random aggregate. The aggregate is then heated to a temperature of $k_BT=0.2\epsilon$ and slowly cooled to $k_BT=0.01\epsilon$ over 250,000 time steps.  Note that this large  dynamic range  in "temperature" is experimentally implemented by variation of interparticle binding free energy (e.g. due to high temperature sensitivity of DNA hybridization).  The annealing cycle is repeated 10 times, with the maximum temperature progressively lowered by $k_B\Delta T=0.01\epsilon$ with each cycle. This is represented in Fig.~\ref{fig:annealing}. This results in a periodic temperature spiking which allows the system to overcome poor configurations by knocking the system out of any local minima in the energy landscape. 

\subsection{\label{sec:Analysis_Methods} Characterization}

The organization of particles upon their self-assembly may be characterized by a number of local order parameters. In particular, bond orientation order parameters (BOPs) are commonly used to identify local ordering \cite{Nelson_2D,Nelson_PRB,Nelson_Toner,Steinhardt}. That approach however has its limitations as it relies on rotationally-invariant (scalar) descriptors of the structure which are not symmetry-specific. Since our system contains anisotropic PPs, a more sensitive characterization method may be utilized. In particular, the relative alignment of both Oh-PPs and Sq-PPs may be captured by a cubatic order parameter \cite{Torq_superballs2010,cubatic_tensor}, the 4$^{\mathrm{th}}$-order tensor that generalizes the conventional nematic order parameter for the cubic symmetry case. More generally, this type of ordering can be characterized by Polyhedral Nematic Order Parameters (PNOPs), which are  traceless, symmetric tensors of $l^{\mathrm{th}}$-order,  $\mathbf{\hat{S}}^{(l)}$ \cite{Torq_superballs2010, cubatic_tensor, Akbari_Glotzer_2015, polyhedral_PRE2016, polyhedral_PRE2016, polyhedral_PRE2018}. In the 3D case, these tensor order parameters may be equivalently  represented  as  vectors composed of $2l+1$ spherical harmonics of order $l$: $|s)_l\equiv \{s^{(l,m)}_{m=-l,...,l} \}$. We will employ this representation with the normalization such that $(s_i|s_j)_l=1$ corresponds to a perfect alignment of Oh-PPs `$i$' and `$j$'.

\begin{figure}[h]
    \centering
    \includegraphics[width=1\columnwidth]{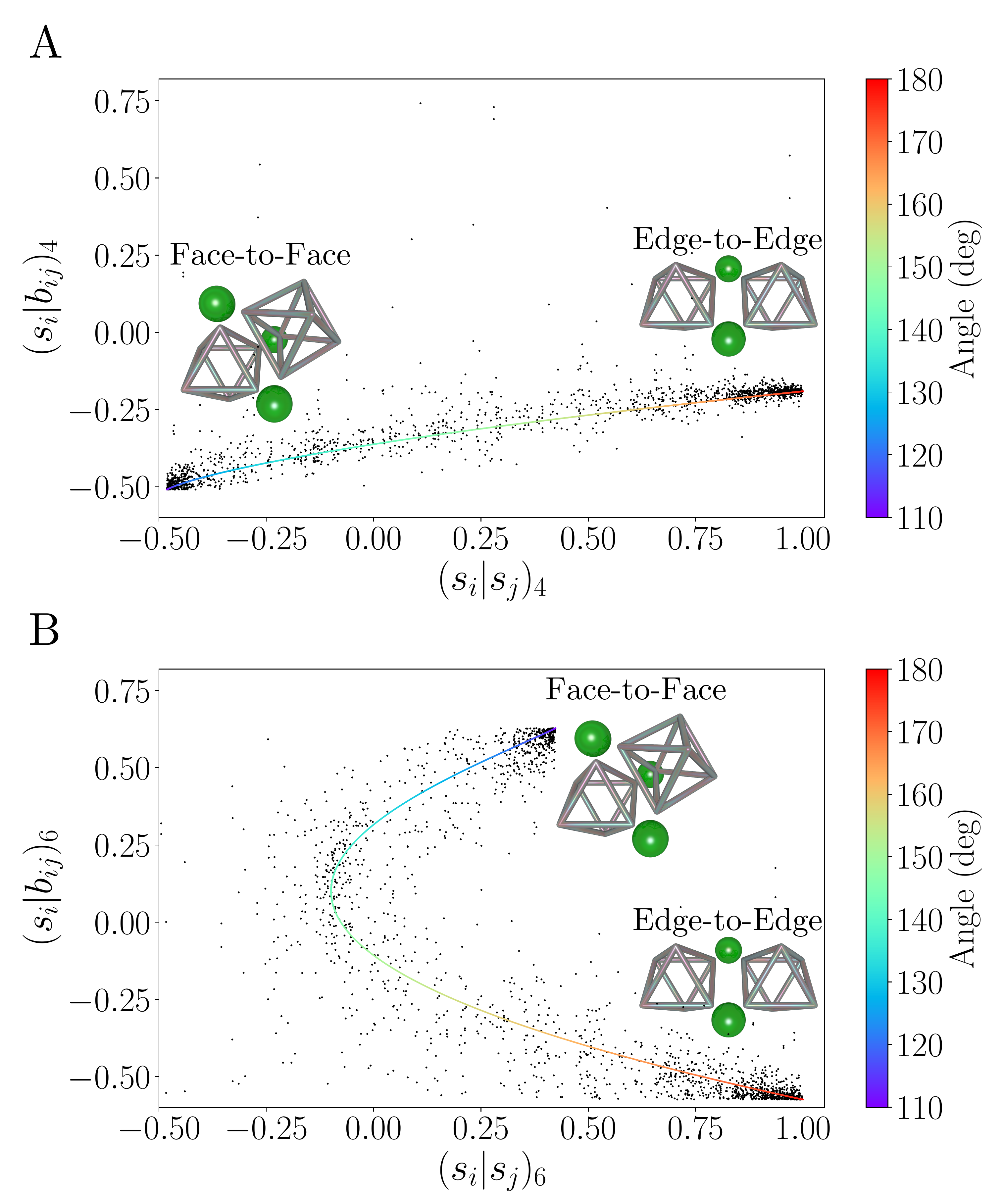}
    \caption{Local order parameter scatter measurements for size ratio 2.08. Each point represents a single bond in the sample. Two clusters of points are seen that correspond to face-to-face and edge-to-edge bonding on the left and right, respectively. Examples of the types of bonds are shown as insets above their respective clusters.}
    \label{fig:SymBOP_plot}
\end{figure}

Furthermore, the  anisotropy of the building blocks enables us to use symmetrized bond order parameters, SymBOPs, that are introduced and discussed in greater detail in Ref.~\onlinecite{symbop}. Specifically, the SymBOP may be defined for any bond, i.e., pair of neighboring particles $i$ and $j$, by projecting the spherical harmonic associated with the bond direction, $\left|b_{ij}\right)_l$ onto the PNOP of one of the particles involved: $(s_i|b_{ij})_l$. In other words, rather than using generic rotationally invariant BOPs as descriptors of the local crystallinity, we detect local order with respect to the natural coordinate system of a particle. In this study, motivated by the geometry of the PPs, we use PNOPs $|s)_l$ with cubic symmetry, for $l=4$ and $6$.  After individual bonds  consistent with the specific order are identified, a bond percolation procedure is employed to find coherent domains with that morphology.

\onecolumngrid
 
\begin{figure}[h!]
    \centering
     \includegraphics[width=1\columnwidth]{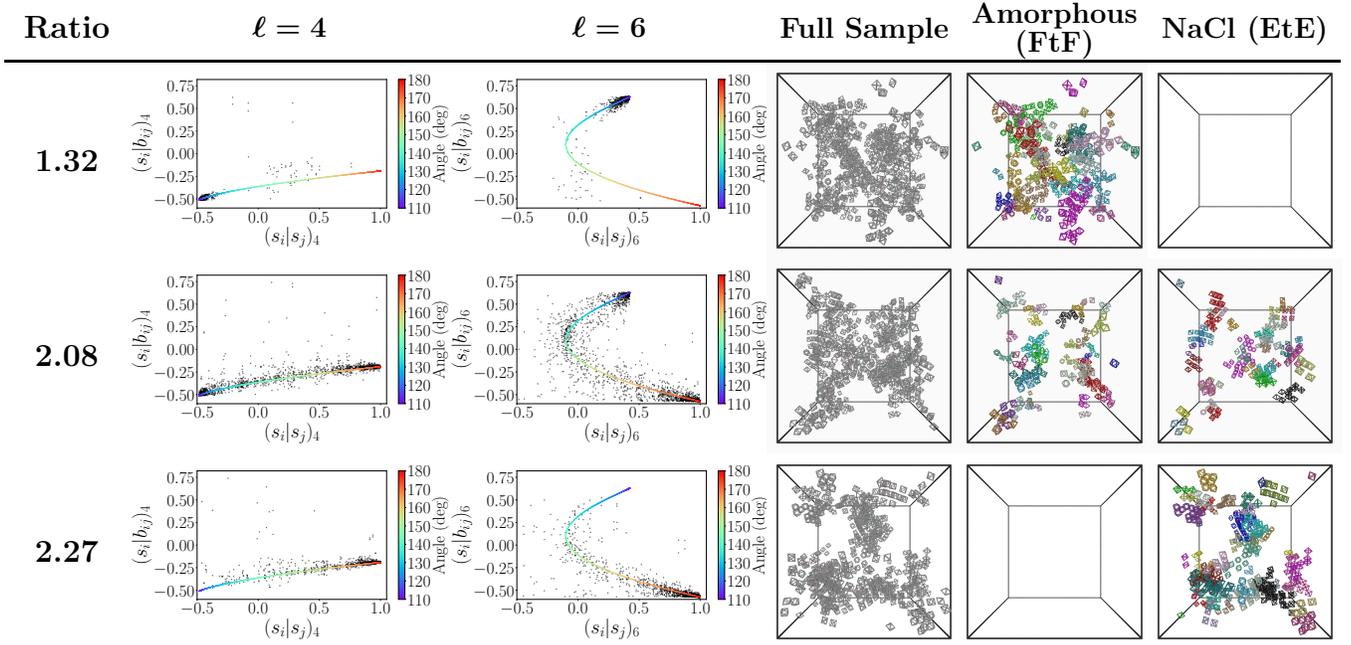}
    \caption{A broad look at how the phase depends on size ratio in IP/Oh-PP systems. The local PNOP/SymBOP scatter plots ($l=4,6$) for Oh-PPs (left columns), and visualizations of the domains observed for the corresponding ratio (right columns). Each color represents a distinct domain. A single cluster of points located in the FtF region is found for small ratios. As the ratio increases, the points move toward the EtE region. For ratios around 2.00 coexistence is observed between the FtF  amorphous and NaCl phases.}
    \label{fig:octa_results_table}
\end{figure}

\twocolumngrid

\begin{figure}[h]
    \centering
     \includegraphics[width=0.9\columnwidth]{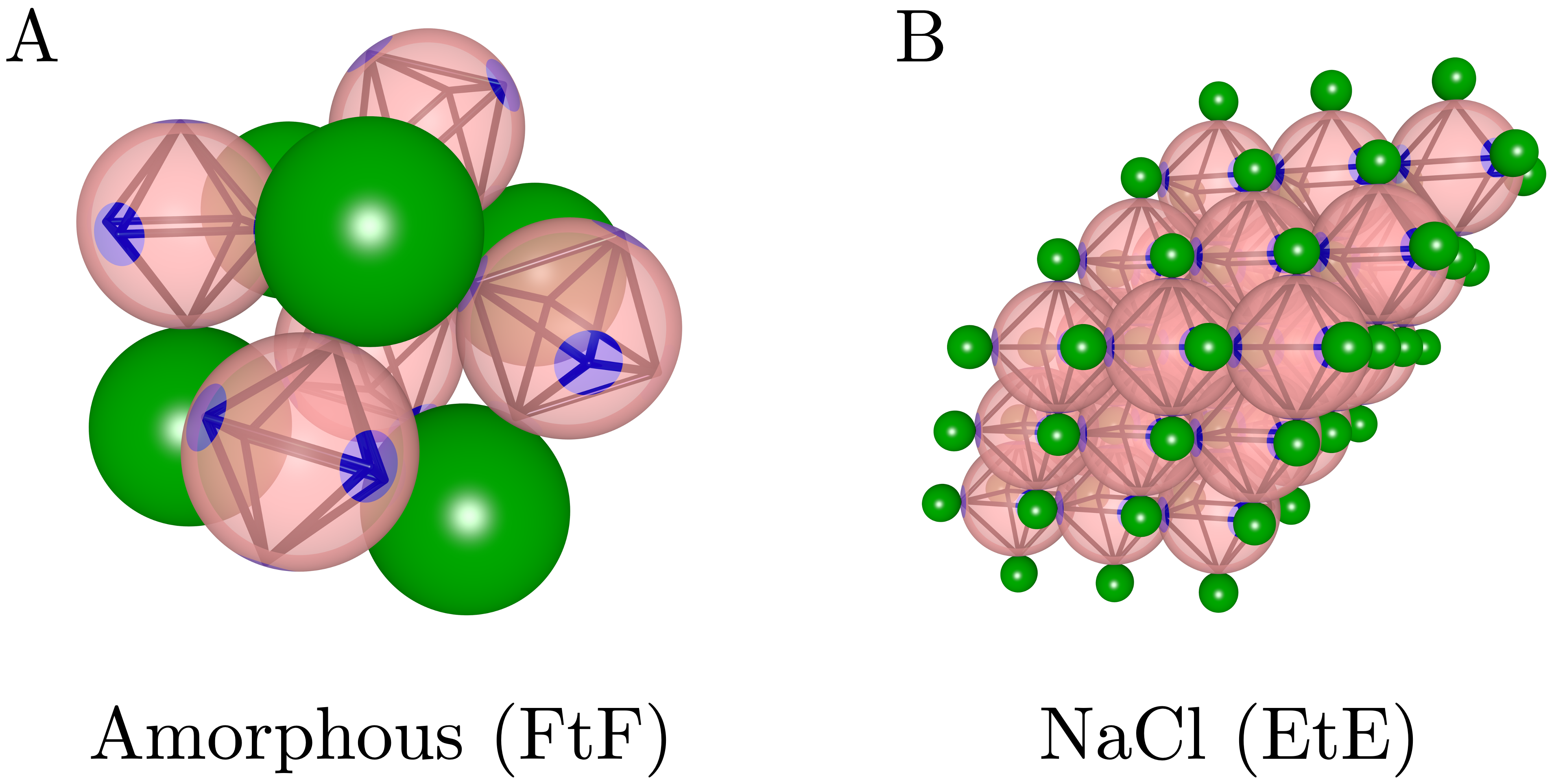}
    \caption{Ideal domains of the structures identified for the IP/Oh-PP system. (A) The FtF phase is amorphous and does not form an extended lattice with long-range order. (B) The EtE phase has long-range order and is found to form a NaCl-type lattice.}
    \label{fig:octa_ideal_lattices}
\end{figure}
\begin{figure}[h]
    \centering
     \includegraphics[width=1\columnwidth]{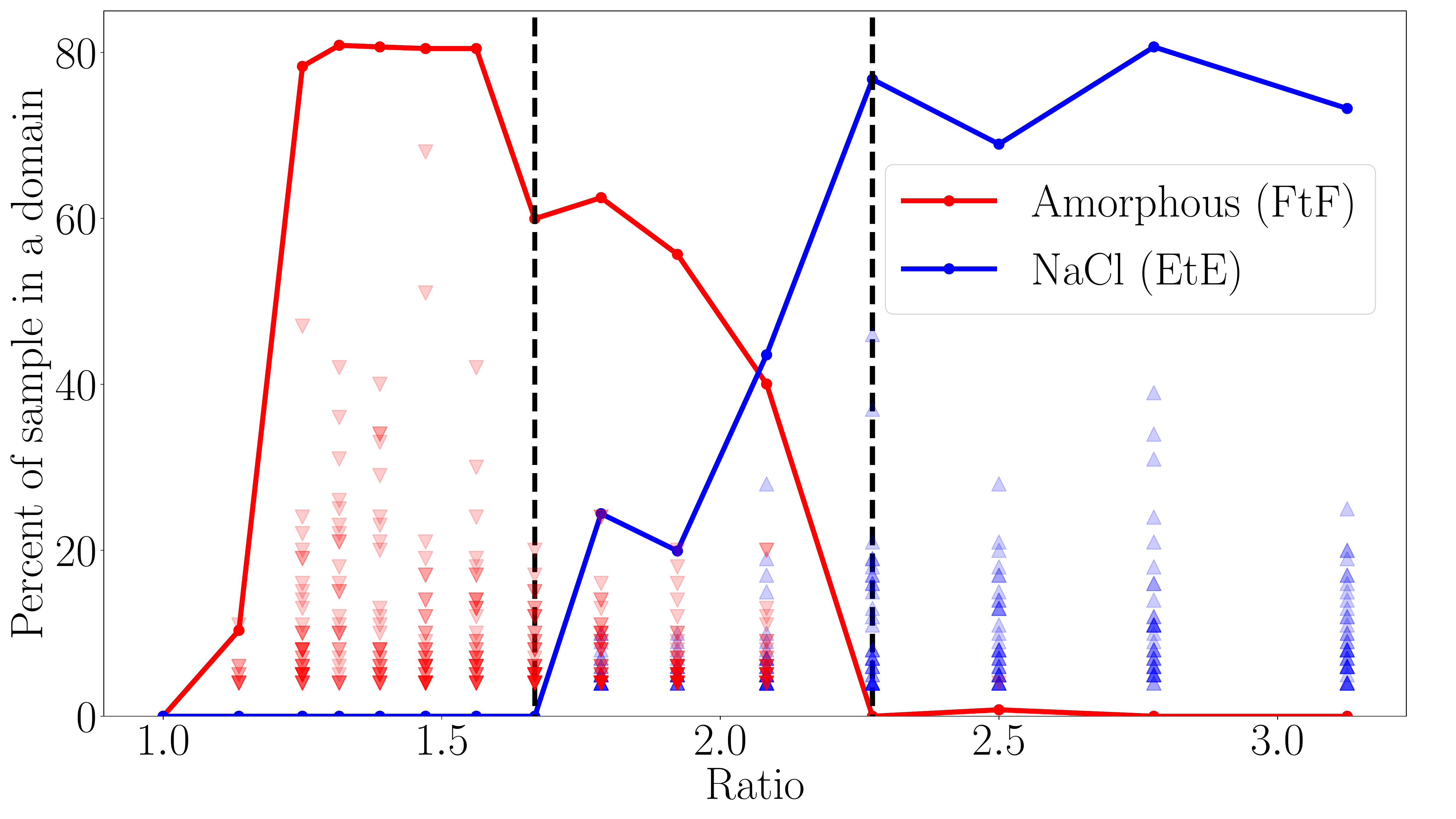}
    \caption{Domain distribution for IP/Oh-PP systems. The lines show the percentage of a sample that was found to be in a domain. The triangles show the sizes of domains found for each size ratio of PP to IP. The dashed vertical lines mark the coexistence region between the two phases.}
    \label{fig:octa_percent_of_sample_in_a_domain}
\end{figure}

\onecolumngrid

\begin{figure}[h]
    \centering
     \includegraphics[width=1\columnwidth]{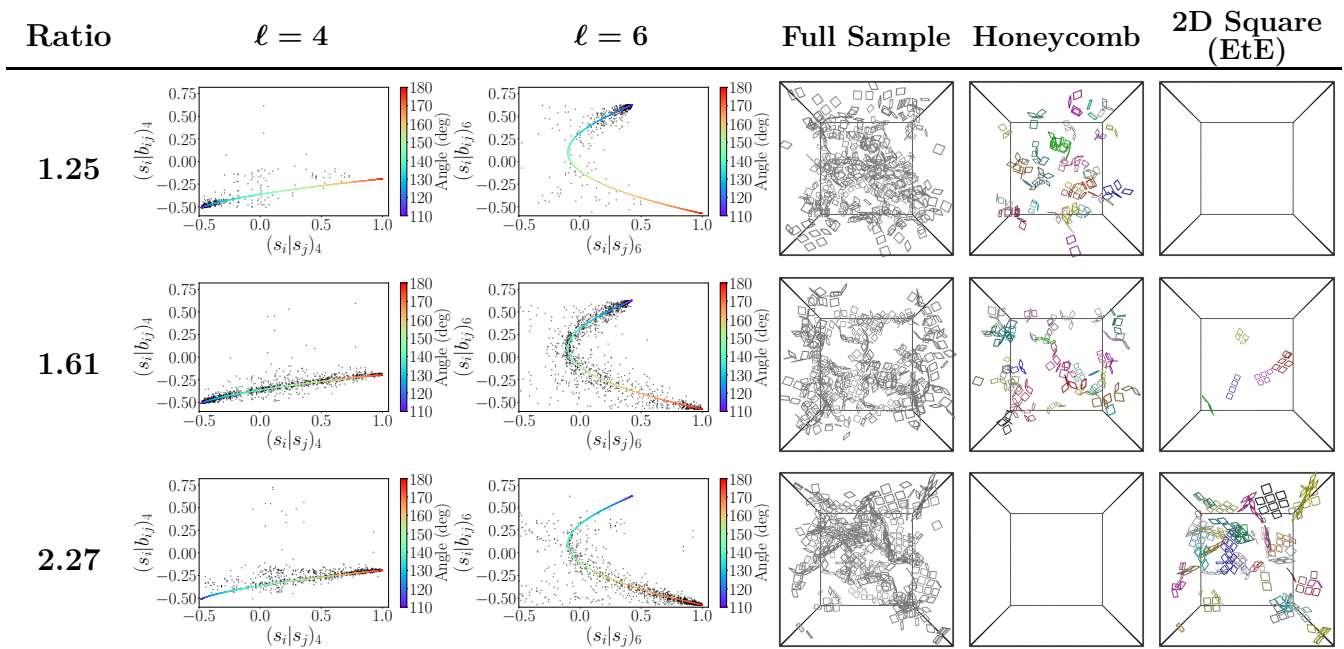}
    \caption{A broad look at how the phase depends on size ratio in IP/Sq-PP systems. The local PNOP/SymBOP scatter plots ($l=4,6$) for Sq-PPs (left columns), and visualizations of the domains observed for the corresponding ratio (right columns). Each color represents a distinct domain. A single cluster of points located in the FtF region is found for small ratios. As the ratio increases, the points move toward the EtE region. For ratios around 1.60 coexistence is observed between the honeycomb and 2D square phases.}
    \label{fig:sq_results_table}
\end{figure}

\twocolumngrid

\begin{figure}[h]
    \centering
     \includegraphics[width=1\columnwidth]{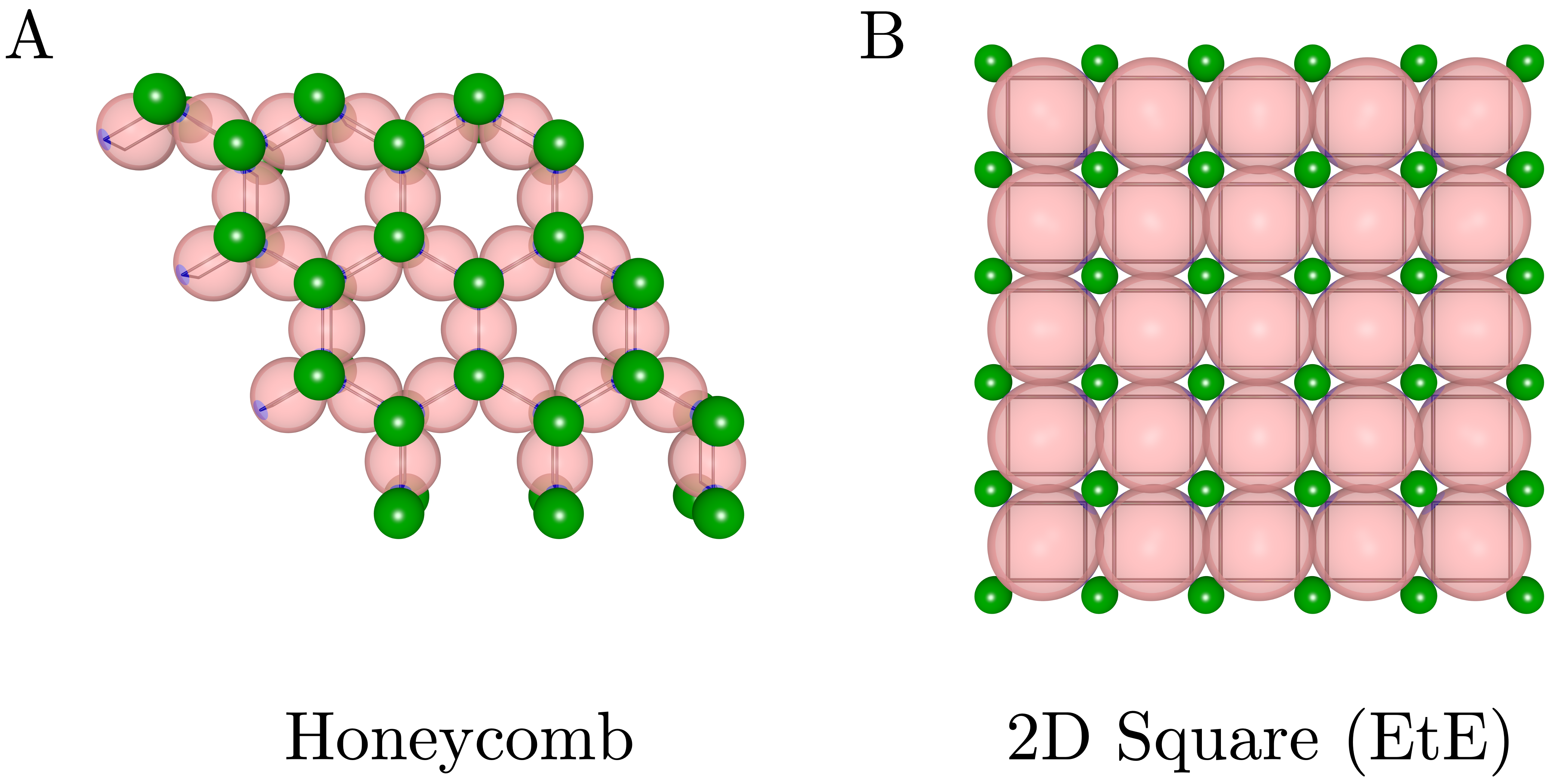}
    \caption{Ideal domains of the order found for each phase of the Sq-PPs: (A) The 3D honeycomb stacking; (B) The EtE phase has long-range order and is found to form a 2D square lattice.}
    \label{fig:sq_ideal_lattices}
\end{figure}

\begin{figure}[h]
    \centering
     \includegraphics[width=1\columnwidth]{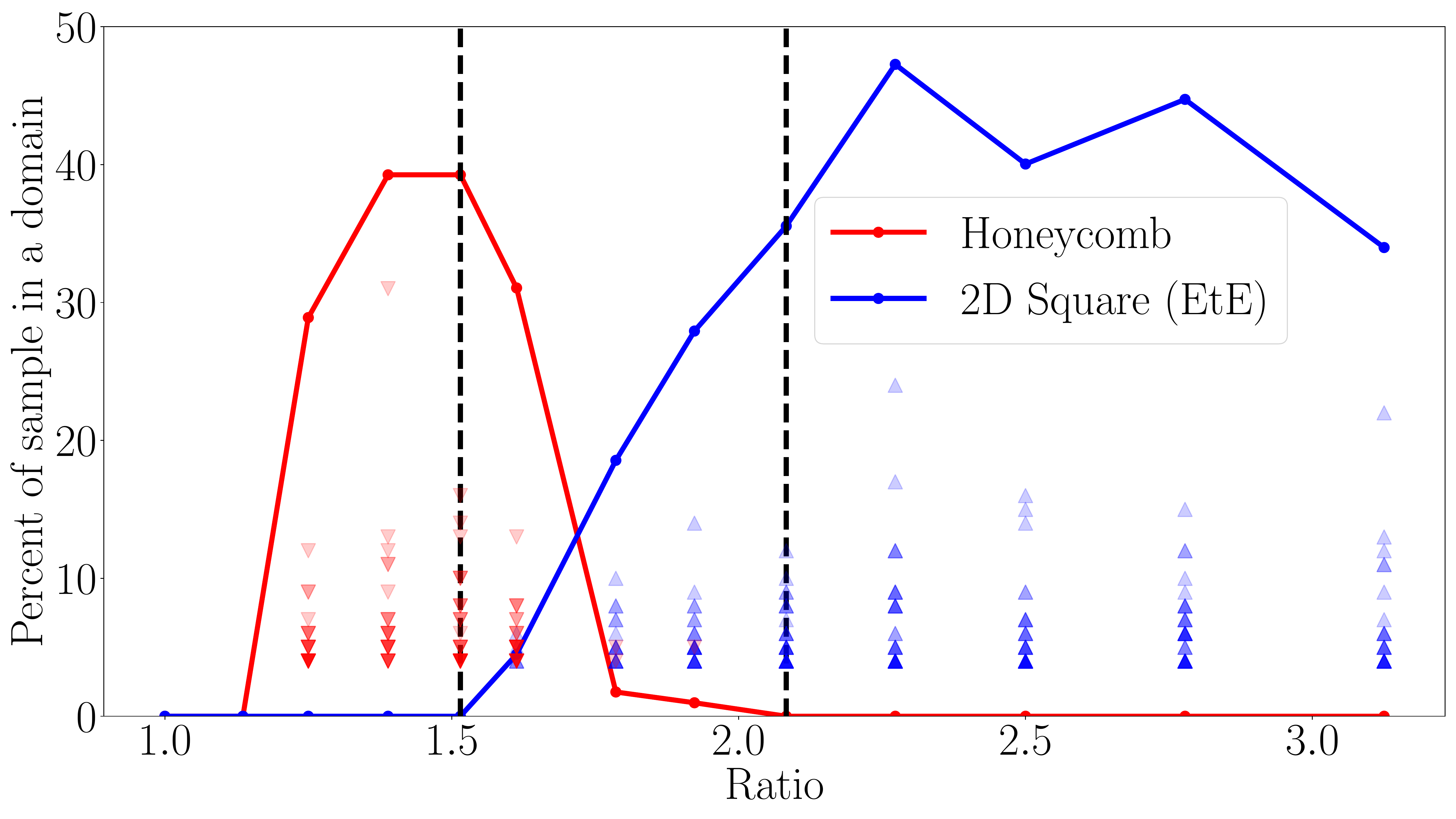}
    \caption{Domain distribution for Sq-PP systems. The colored solid lines show the percentage of a sample that was found to be in a domain. The triangles show the sizes of domains found for each size ratio of PP to IP. The dashed vertical lines mark the coexistence region between the two phases.}
    \label{fig:sq_percent_of_sample_in_a_domain}
\end{figure}

\section{\label{sec:Results} Results}

The self-assembly of   1:1 IP/PP mixtures are   simulated,  for  particle  size ratios, $\sigma_{PP}/\sigma_{IP}$, ranging from 1.00 to 3.13.
Figure~\ref{fig:SymBOP_plot} shows a "portrait" of a typical self-assembled structure as a scatter plots in the PNOP/SymBOP space. The two  coordinates represent the relative alignment of the two neighboring particles,  $(s_i|s_j)_l$, and the SymBOP associated with their bond, $(s_i|b_{ij})_l$, respectively (for $l=4,6$). Each bond is represented by two points, since  $|b_{ij})_l$   can be projected onto the reference vector of any of the two particles involved, $|s_i)_l$  and $|s_j)_l$.      
Two distinct clusters of bonds are observed. A color-coded curve follows the transition in the bonds from edge-to-edge (EtE) on the right (red) to face-to-face (FtF) on the left (blue). The curve is constructed by beginning with two Oh-PPs sharing two IPs in an EtE configuration, with all four particles lying in a common plane, as shown in the inset on the right of Fig.~\ref{fig:SymBOP_plot}A. The rotation of one Oh-PP about the IPs to the FtF configuration (left inset) traces the curve from red to blue. The color bar corresponds to the angle formed from the center of one Oh-PP, to the point between the IPs, to the center of the second Oh-PP. For Oh-PPs, FtF corresponds to a special type of amorphous structure, while EtE binding signals emergence of ordered domains, i.e. NaCl lattice. For squares, the analog of FtF represents fragments of 3D honeycomb stacking, while EtE binding gives rise to domains that arrange in a two-dimensional square lattice. Below we present results for Oh-PP and Sq-PP systems, respectively.

\subsection{\label{sec:results_for_octahedra} Self-assembly of  IP/Oh-PP hybrid system}

Fig.~\ref{fig:octa_results_table} represents selected  results  for IP/Oh-PPs system. Each row represents a single size ratio. It shows PNOP/SymBOP portraits of the self-assembled structure, discussed above, alongside the  real-space images, both without and with individual domains identified. 
For size ratio below  1.00, the self assembly  does not yield any ordered domains. This result is expected as the interaction vanishes exactly at the distance the IPs would bond to a patch (see Eq.~\ref{eq:soft_pot}). The attraction between the IPs and the patches is stronger for ratios greater than 1.00. However, for small ratios the attraction is weak and the IPs only feel the shallow tail of the potential. Despite this, the incipience of bond clustering in PNOP/SymBOP plot is observed around a ratio of 1.14. As mentioned above, the type of bonding for small ratios of Oh-PPs is FtF. An ideal domain of this type is seen in Fig.~\ref{fig:octa_ideal_lattices}A. Although the domains made solely from this type of bond can be large (nearly 70 PPs for a ratio of 1.47), they are amorphous, i.e. not space filling and  and don't have any long-range order.

For a given ratio, the solid red curve in Fig.~\ref{fig:octa_percent_of_sample_in_a_domain} shows the percentage of the final state that was found to be in some domain, and the red semi-transparent arrows use the same vertical axis to list all of the domain sizes observed. For ratio 1.14, only about 10\% of the whole sample is found to be in an ordered domain, and on average the domains have a size of about 5 PPs each. 

Increasing the ratio nominally to around 1.25 results in a jump in both the number of ordered domains and the distribution of domain sizes. The size ratios ranging from 1.25 to 1.56 show nearly 80\% of the final state to be in some domain. Above 1.56 the prevalence of the amorphous domains begin to subside, and a second type of structure is observed. The new structure has an EtE bonding and a high degree of orientational ordering between constituent particles. The EtE bonded domains possess long-range order and form a NaCl lattice, as shown in Fig.~\ref{fig:octa_ideal_lattices}B.

Between the size ratios of 1.67 and 2.27, coexistence of the two types of domains is observed. As seen in Fig.~\ref{fig:octa_percent_of_sample_in_a_domain}, the decrease in particles in amorphous domains is completely compensated by the increase in particles in the NaCl domains. Above a ratio of 2.08 the amorphous phase vanishes and the NaCl phase dominates. The simulations were performed up to a ratio of 3.13, with the NaCl domains continuing to persist.

The transition from amorphous (FtF) to NaCl (EtE) is clearly seen in Fig.~\ref{fig:octa_results_table}. The two local order parameter scatter plots ($l=4,6$) on the left demonstrate that points cluster only around the FtF region for smaller ratios ($\sim$ 1.14 to 1.67), as was shown in Fig.~\ref{fig:SymBOP_plot}. As the ratio increases, the points move away from the FtF region of the scatter plot toward EtE. Coexistence between the two phases at a ratio of 2.08 is evident from Fig.~\ref{fig:octa_results_table}. The locations of the clusters agree with the domains shown to the right of the plots. The last row corresponding to ratio 2.27 shows only the long-range ordered NaCl phase to remain.

\subsection% {\label{sec:results_for_squares}
{Self-assembly  of  IP/Sq-PP hybrid system}

The analysis of the systems with Sq-PPs is analogous to that corresponding to the systems with the Oh-PPs. This approach introduces consistency in the characterization and simplifies the explanation of the types of bonding observed through the local order parameter calculations. 

Sq-PPs show a similar trend as that for the Oh-PPs, namely that the smallest size ratios do not demonstrate order. The first domains begin to appear for a size ratio of 1.25. A comparison of the lowest ratio order parameter calculations (in Fig.~\ref{fig:sq_results_table}) for Sq-PPs to Oh-PPs (in Fig.~\ref{fig:octa_results_table}) demonstrates the advantage of using the same octahedrally-symmetric order parameters for Sq-PP. Both plots show distinct clustering of points near what is determined to be FtF bonding for Oh-PPs. While it cannot be true FtF bonding for Sq-PPs, the color bar for the plot shows that tangents to the surfaces of each square come together at an angle of $110^\circ$ to $120^\circ$. Most of the identified domains have a zig-zag shape,  consistent with being fragments of a 3D honeycomb stacking lattice. Figure~\ref{fig:sq_ideal_lattices}A shows an ideal version of such a lattice made of IP/Sq-PPs. 

The honeycomb phase peaks for size ratios in the range of 1.39 to 1.52 with about 40\% of the PPs found in some domain and a maximum domain size greater than 30 PPs (see Fig.~\ref{fig:sq_percent_of_sample_in_a_domain}). The larger domains tend to deviate from the honeycomb lattice and become more amorphous. Beyond 1.52, the honeycomb phase begins to diminish and a higher polyhedral nematic order is favored indicating EtE bonding. This type of bonding presents itself as a two dimensional square lattice for Sq-PPs, as seen in Fig.~\ref{fig:sq_ideal_lattices}B. This marks the beginning of a two-phase coexistence between the honeycomb domains and the emerging square lattice domains. The coexistence is short-lived as the honeycomb phase almost completely vanishes by a ratio of 1.79. However, a few small domains are observed to remain with a size ratio up to 1.92. Above this ratio only the square lattice phase is observed. This phase peaks at a ratio 2.27, and starts decreasing at a ratio of 3.13. Note that ideal honeycomb structure would have 2:3 IP/PP composition. However, our simulations of asymmetric mixtures resulted in a much lower yields of either of the structures observed (see Appendix A).

\section{\label{sec:Discussion} Discussion}
%(note: Square FtF hard sphere model movie is uploaded as ``FtF\_sq\_hexagonal\_movie.mp4") \\ 

Both Oh-PPs and Sq-PPs favor FtF bonding for small size ratios and undergo a structural transition to EtE bonding for large ratios. A simple geometrical model elucidates the origin of this transition.

\begin{figure}[h!]
    \centering
    \includegraphics[width=1\columnwidth]{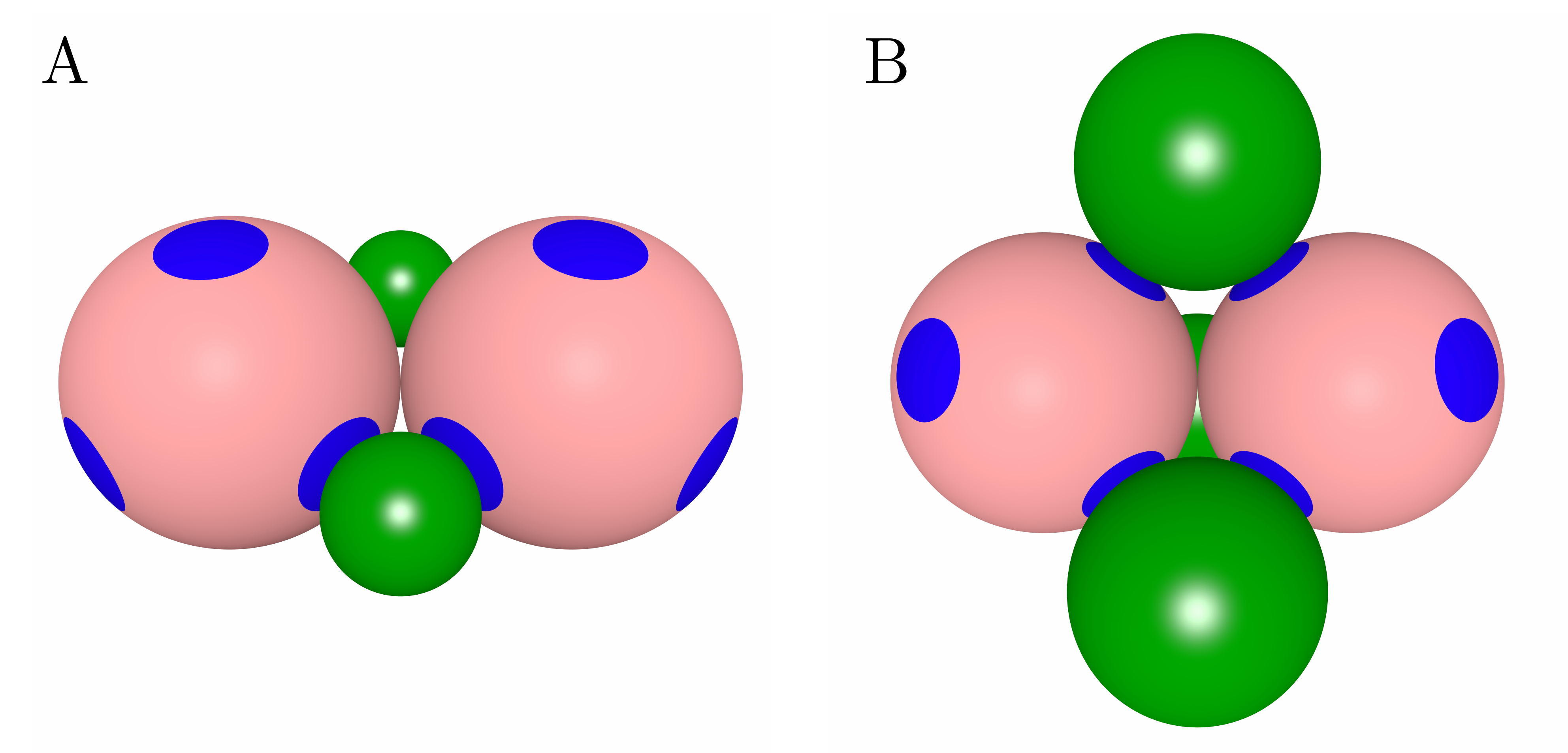}
    \caption{Oh-PP in the (A) NaCl (EtE) configuration and (B) amorphous (FtF) configuration.}
    \label{fig:octa_hard_sphere_geometry}
\end{figure}

% \subsection{Octahedra}

\begin{figure}[h!]
    \centering
    \includegraphics[width=1\columnwidth]{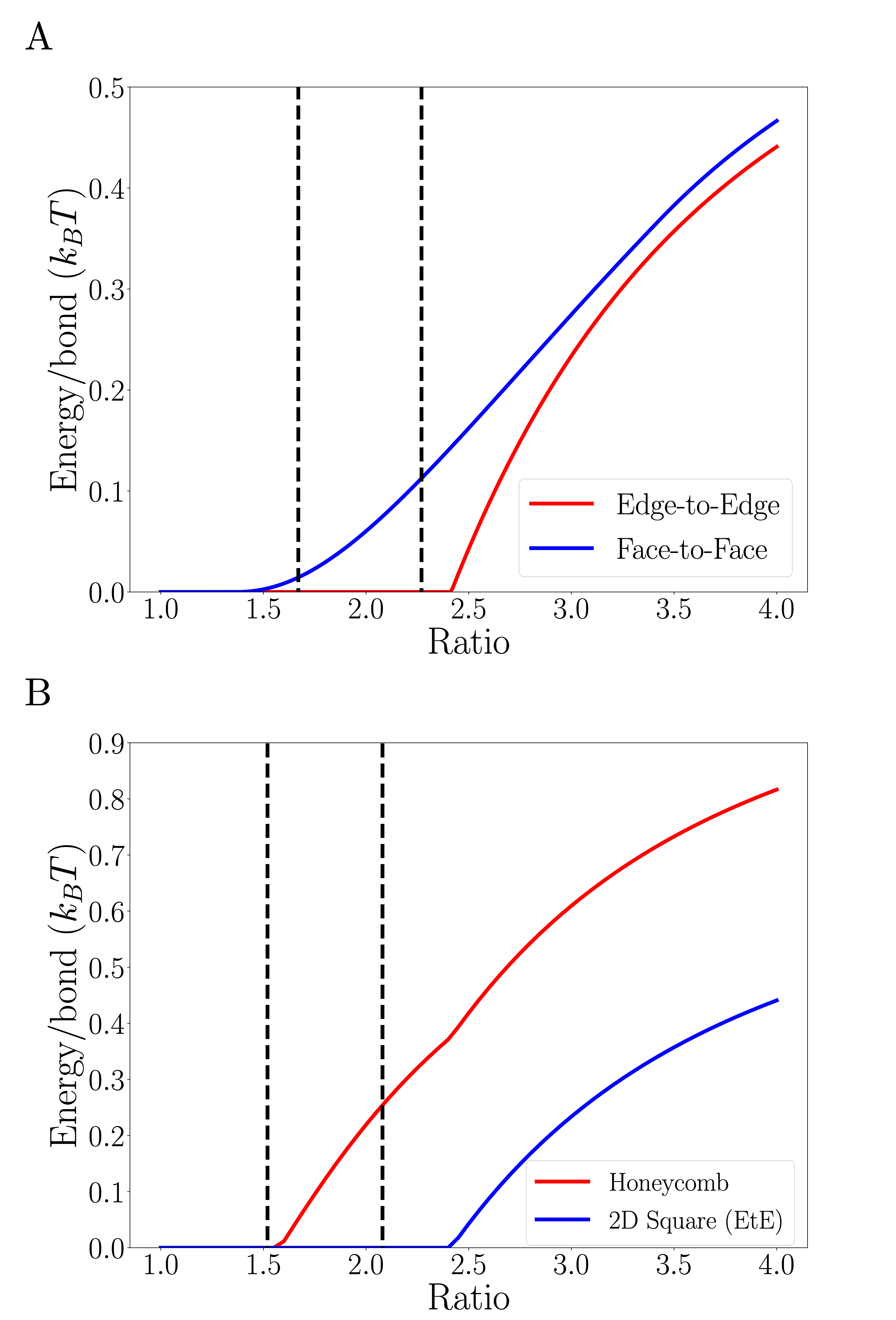}
    \caption{Energy per bond as a function of size ratio for (A) NaCl (EtE) bonding (blue) and amorphous (FtF) bonding (red) and (B) 2D square lattice (EtE) bonding (blue) and honeycomb lattice (red). The dashed black lines mark the two-phase coexistence regions for each type of PP. Above the dashed lines FtF bonding is no longer observed.}
    \label{fig:octa_sq_bonding_energy_HS}
\end{figure}

Consider an Oh-PP that has captured two IPs and is found to be in an EtE configuration, as shown in Fig.~\ref{fig:octa_hard_sphere_geometry}A. For small size ratios, it is energetically favorable for the Oh-PPs to capture a third IP, completing the transition to FtF bonding as shown in Fig.~\ref{fig:octa_hard_sphere_geometry}B. Even if EtE binding corresponds to thermodynamically more stable  structure (wich is likely the case for NaCl lattice),  FtF binding will win kinetically.    As the ratio increases, however, the size of the IP decreases and the PPs must move closer together to keep the IPs bound. At a critical ratio, the PPs make contact and can no longer move closer to preserve the direct contact between the IPs and the patches. The critical ratio marks the point where larger ratios incur an energetic cost to form FtF bonds and lead to the possibility of other structures. Figure~\ref{fig:octa_sq_bonding_energy_HS}A shows the energy per bond for FtF and EtE bonding for different ratios. The zero in energy corresponds to the direct contact bonding between an IP and a patch for all ratios. The plot shows that FtF bonding is initially energetically favorable but begins to encounter an energy barrier for a size ratio around 1.37. The dashed black lines mark the coexistence region of the two structures from BD simulations. The FtF bonding disappears for a size ratio of 2.08, marked by the second dashed line. For higher size ratios, only structures involving EtE bonding are formed. Our interpretation is that the growing energy barrier associated with formation of FtF  amorphous structure promotes EtE binding and the ultimate formation of the ordered domains. 

\begin{figure}[h!]
    \centering
    \includegraphics[width=1\columnwidth]{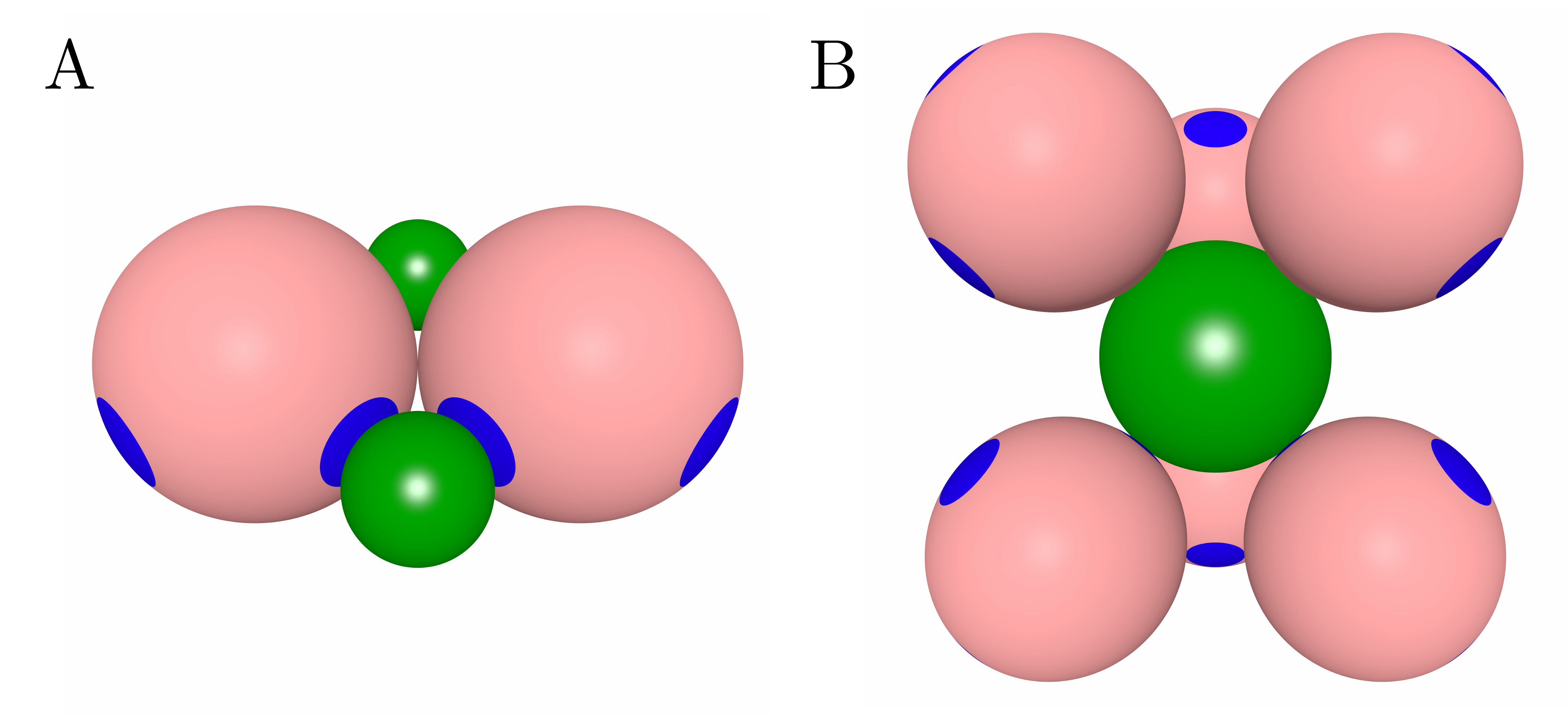}
    \caption{Sq-PP in the (A) 2D Square lattice (EtE) configuration and (B) honeycomb configuration.}
    \label{fig:sq_hard_sphere_geometry}
\end{figure}

% \subsection{Squares}

For Sq-PPs a similar trend is observed; small ratios tend to form a FtF honeycomb structure, while large ratios lead to EtE square lattice domains. In a FtF domain, six PPs share a single IP. Considering the hard sphere model shown in Fig.~\ref{fig:sq_hard_sphere_geometry}. As the ratio increases, the PPs must move closer to stay bonded with the IP. This leads to two critical ratios. The first ratio is encountered when the PPs within a single layer make contact at a size ratio of around 1.58. Above this ratio the IP begins to move away from the patches of the PPs, initiating the initial divergence from the minimal energy observed in Fig.~\ref{fig:octa_sq_bonding_energy_HS}B. At this point, the IP cannot be kept directly on the patches of each PP, but the PP layers still move closer together as the size ratio continues to increase. The energy cost effected above this critical ratio leads to the emergence of the EtE structure, which appears at a ratio of 1.61 and remains through 3.13. The dashed black lines mark the coexistence region of EtE and FtF bonds. Above the ratio of 1.92, the energy cost is too great and the only stable structure formed is via EtE bonds.

\section{Conclusion}
We studied self-assembly in hybrid IP/PP systems, with two arrangements of patches: octahedral and square.  The assembly was additionally facilitated by the annealing procedure.   The anisotropic nature of the particles enabled the use of PNOPs, together with a new characterization technique\cite{symbop}, based on  SymBOPs.  The latter generalizes the classical concept of bond orientation order parameter, and allows  to look for distinct morphological domains, both ordered and amorphous,   within the assembled structures. Using the PNOP/SymBOP analysis, the phase behavior  for both types of PPs have been studied, as a function of the PP/IP size ratio. Importantly, our focus was not on determining the equilibrium phase diagram, but to identify structures that can self-assemble when kinetic limitations are taken into account.  Both types of PPs were observed to exhibit two phases over size ratios from 1.00 to 3.13.
% , with coexistence occuring for ratios  around 1.67 to 2.27, and 1.52 to 2.08, for Oh-PP and Sq-PP, respectively. 

The higher symmetry phases, NaCl and square lattice, respectively, are observed above size ratio $2$. Below the size ratio $1.5$, the dominant morphology of Oh-PP/IP system is a highly compact  structure with ``Face-to-Face" arrangement of the anisotropic particles. While lacking long range order, this amorphous phase is substantially different from a generic random aggregate. In particular, it has a pronounced PNOP/SymBOP signature that allowed us to clearly identify its individual domains, a rather unusual feature for  a disordered structure. In the case of Sq-PPs, the lower size ratio  corresponds to another peculiar partially ordered structure. It features ziz-zag-like patterns, that are likely to be  fragments or precursors of a 3D honeycomb stacking. Once again, the domains are easily identifiable with the help of the PNOP/SymBOP characterization technique.  This demonstrates, on the one hand, the potential of the hybrid PP/IP platform for self-assembly of non-trivial morphologies, and, on the other hand, the power of the PNOP/SymBOP-based characterization.  All code and run input files for simulations are available on GitHub \url{https://github.com/duttm/Octahedra_Nanoparticle_Project}.

\begin{acknowledgments}
This research was partially done at, and used
resources of the Center for Functional Nanomaterials, which is a U.S.
DOE Office of Science Facility, at Brookhaven National Laboratory under
Contract No.~DE-SC0012704. M.D. ackowledges financial support from NSF CAREER award DMR-1654325. S.C.M. acknowledges financial support from NSF award OAC-1547580 and the Chemical and Biochemical Engineering Department at Rutgers.
\end{acknowledgments}

\appendix

\renewcommand{\thefigure}{A\arabic{figure}}
\setcounter{figure}{0}
\section{Sq-PPs for different compositions}

\begin{figure}[h!]
    \centering
     \includegraphics[width=1\columnwidth]{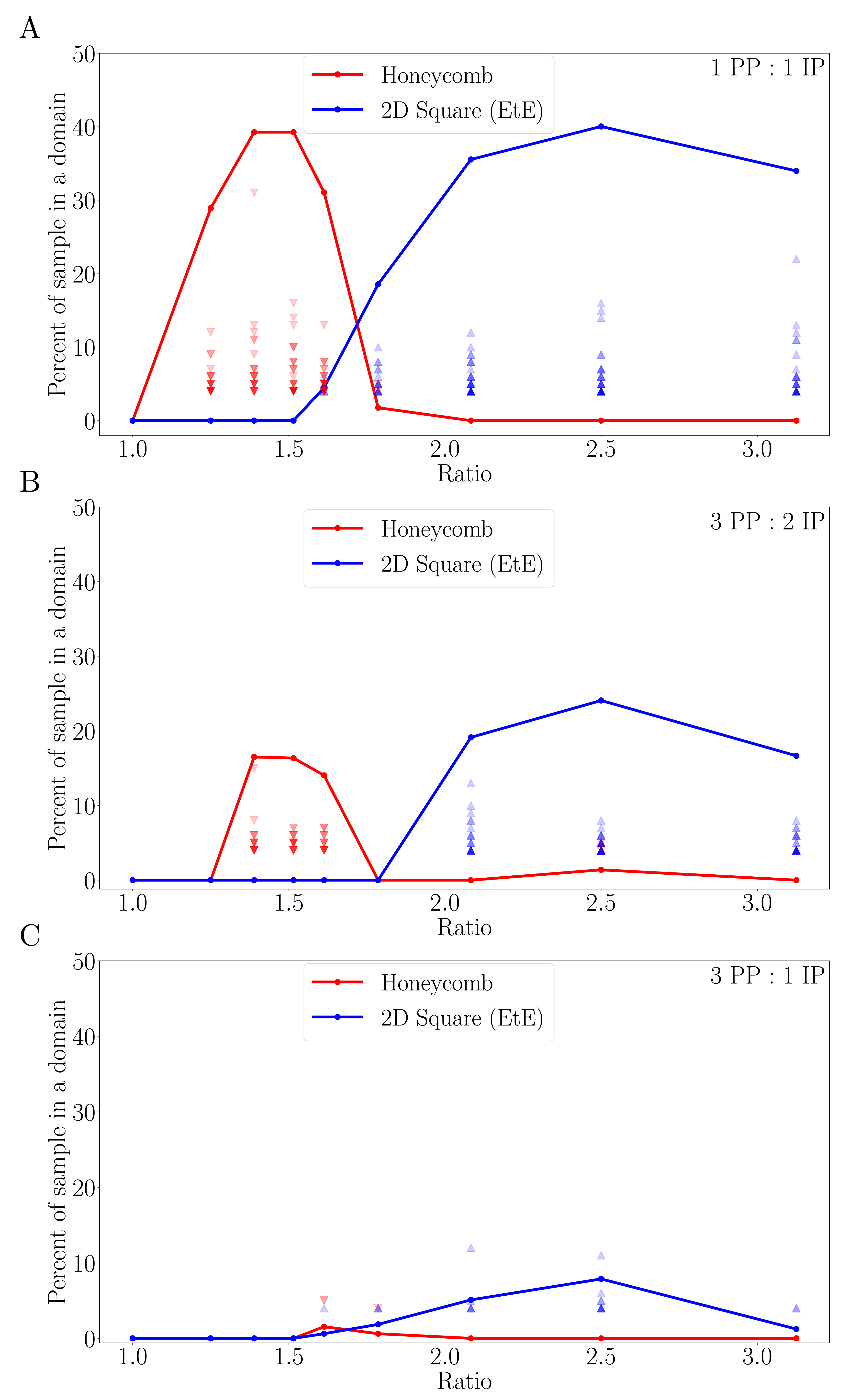}
    \caption{Plots displaying the change in observed domains with a change in composition. The colored solid lines show the percentage of PPs that are found in a domain for the honeycomb (red) and the 2D square lattice (blue). The composition begins at (A) 1 PP : 1 IP; the same composition used for the simulations in the main text. As the IPs become less abundant such as in (B) 3 PP : 2 IP and (C) 3 PP : 1 IP, the amount of order decreases.}
    \label{fig:sq_plots_by_composition}
\end{figure}

Since ideal honeycomb stacking has 3:2 PP/IP composition,  investigated how the composition of the mixture affects the self-assembly in IP/Sq-PPs system. The simulations discussed in the main text had a composition of one PP to one IP. In addition to this, we looked at compositions 3:2 and 3:1 for select PP to IP size ratios. The results for different compositions can be seen in Fig.~\ref{fig:sq_plots_by_composition}. 

\bibliography{main.bib}

\end{document}